\documentstyle[12pt]{article}
\begin{document}

\title{ENERGETICS OF THE SUPERFLARE FROM SGR1806-20 AND A POSSIBLE
ASSOCIATED GRAVITATIONAL WAVE BURST}

\author{J. E. HORVATH$^*$\\
Instituto de Astronomia, Geof\'\i sica e Ciencias Atmosf\'ericas\\
Rua do
Mat\~ao 1226, 05508-900 S\~ao Paulo SP, Brazil\\
$^*$foton@astro.iag.usp.br\\}

\maketitle

\abstract{ We discuss in this Letter the energetics of large
gamma-ray superflares observed from Soft-Gamma Repeater sources.
The last recorded event has in fact ruled out some models for the
energy release. For the first time actual information about a
possible associated gravitational wave emission may be gathered
from the LIGO data, even in the case that most of the energy was
emitted in gamma-rays. Even upper limits on the amplitude of the
latter $h$ at the expected frequency $f_{c} \, \leq \, 2 \, kHz$
may be useful to further constrain the remaining mechanisms}

\section{Introduction}

A small class of compact objects, the so-called Soft Gamma
Repeaters, has recently attracted the attention of astrophysicists
again. The few members of the class are now thought to be highly
magnetized neutron stars (or magnetars, \cite{TD}), in spite that
not all observed features fit so easily in this picture
\cite{Lin}. In steady state SGRs emit small quasi-thermal bursts
with an associated temperature of $\sim \, 10 \, keV$. This steady
phase is probably of short duration, since the total number in the
galaxy may not be larger that $\sim\, 5-10$ objects.
Interestingly, a quake-like pattern, similar to the terrestrial
earthquakes, has been found in the time series \cite{Cheng} and
therefore the idea that crust cracking was responsible for the
observed behavior was put forward.

While the steady behavior of these sources is quite interesting,
and not yet fully understood, more spectacular events definitely
associated with them challenge the ingenuity of the researchers.
These are the {\it superflares}, sudden events releasing a large
amount of energy observed to occur once in three out of the four
firmly known sources, namely SGR 0526-66,SGR 1900+14 and SGR
1806-20. The existence and nature of superflares may reveal
something fundamental about these sources. We focus in this note
on the energy budget quest of these events, with possible
implications for the existing gravitational (GW) wave experiments.

\section{Free-energy sources of superflares}

One of the most important and primary issues is the very source of
the released energy (see Table 1,Column 3 for the inferred
isotropic equivalent values). An interpretation in terms of the
magnetar model has been advanced, involving a large-scale
reconnection/untwisting of the magnetosphere. This kind of model
seems to fit the observed features (rise time, spectral features
and decay) quite well \cite{Hur}. However, and even if a high
magnetic field is involved, there are a few alternatives for the
source of free energy which may be worth considering for the sake
of comparison and further study.

The first one is some kind of phase transition at high density.
Depending on parameters, and also on the detailed dynamics in
which matter undergoes a phase transition, a huge amount of energy
may be released simply by a mechanical readjustment of the star
\cite{MVP}. Those calculations arrive at an upper limit to the
released energy of the order of $\Delta E \, = \, W(\delta R/R)$,
where $W$ is the original binding energy of the star and $\delta
R/R$ is the fractional change in the radius. Numbers as high as
$10^{53} \, erg$ seem possible, although not all this released
energy will be funnelled in a definite channel and must be shared
(see below). Another picture \cite{BHV} invokes a more complicated
and exotic structure at high density to produce the superflares
spaced in time, while the ``active" phase results in between.
Admittedly, there is a high degree of speculation in structured
quark matter that has not been settled over the years
\cite{Curt,SPM,Heisel}.

A new version of this picture (an exotic solid quark phase which
occupies most of the interior of the "neutron" star
\cite{Xu1,Xu2}) has been recently proposed as a likely source of
free energy. In these models, condensation of quarks in position
space competes with the celebrated momentum space features (CFL
and analogues). This solid may {\it crack} whenever sufficiently
strained, releasing an elastic energy of the order of $\Delta E \,
= 10^{47} (\mu/10^{32} erg \, cm^{-3}) (\theta_{max}/10^{-3})
(R/10 \, km)^{3} \, erg$, assuming a high value of the shear
modulus as appropriated for the exotic solid \cite{Xu1,Xu2}, a
maximum strain $\theta_{max} \ \sim \, 10^{-3}$ and essentially
all the star as a solid body participating in the process.
Stresses may be caused by spindown, which is unlikely for these
slow rotators with periods $\sim \, seconds$, or by magnetic
fields \cite{Rud,Pache}. A simple estimation of the field
necessary to fulfill the condition of cracking can be made
balancing the magnetic and solid stresses, yielding the lower
limit $ B \, > \, 2.5 \times 10^{15} (\mu/10^{32} erg \,
cm^{-3})(\theta_{max}/10^{-3})(B_{c}/10^{15} \, G)^{-1} \, G$. We
see that, even if a high magnetic field could be ultimately
confirmed, its role in superflares would not be automatically
established. Solid quark stars quakes are good candidates to give
rise to an interesting experimental signal (next Section).

Alternatively, another more exotic kind of free energy may be a
source, namely the "burning" of neutron matter to strange matter
\cite{Bug}. This ultimate source of free energy arises if the
energy per baryon number unit of three-flavor quark matter happens
to be {\it lower} than the same quantity in the confined hadronic
phase, and if further astrophysical conditions are met to trigger
the conversion of a neutron star well after its birth, analogously
to models of GRBs \cite{Bomba}. There is plenty of free energy
available from this process, since the conversion is expected to
release $\sim \, 10 \, MeV$ per baryon number unit, adding up to
$\Delta E \, = 10^{52} (\epsilon/10 \, MeV)(N_{B}/10^{57}) \, erg$
for a solar-mass star. When and how this energy is employed
depends on the details of the process \cite{Bug}. A mechanical
readjustment of the star with release of additional energy may
follow the conversion and play a role \cite{DB}. Needless to say,
the observation of a second superflare from the same source would
rule out this possibility.

From a general point of view we may classify these few sources of
free energy as "mechanical" (first two cases), "chemical" (strange
matter hypothesis), and "external" (magnetospheric hypothesis, see
also Refs. 17 and 18 for an alternative accretion models also
falling in this category ). It should be emphasized that only the
last superflare observed from the source SGR 1806-20 firmly
excludes models based on conventional crust seismology, since the
energy detected in gamma-rays alone exceeds the total elastic
energy of any reasonable model crust. Even if we allow a
substantial beaming factor $\sim \, 0.1$, thus reducing the
released energy, these models are in trouble to explain
superflares, even more if recurrence is ever observed in any of
the sources and particularly the recent SGR1806-20. In other
words, if superflares are due to solid cracking, the latter should
be exotic. The detailed lightcurves can be a powerful tool to
check the details of specific models.

\bigskip

\noindent{\bf Table 1. Superflares from SGRs and detectability of
associated GW}

\begin{table}[h]

{\begin{tabular}{@{}cccccccc@{}}
& & & Object & Date & $E_{\gamma}$ (erg) & D (kpc) & Maximum $\eta$\\
& &    & SGR 0526-66 & 1979 Mar 5  & $6 \times 10^{44}$ & 50 & $2 \times 10^{-3}$\\
& &    & SGR 1900+14 & 1998 Aug 27 & $2 \times 10^{44}$ & 15 & $0.01 $ \\
& &    & SGR 1806-20 & 2004 Dec 27 & $3.5 \times 10^{46}$& 10 & $4.8 $\\
\end{tabular}}
\end{table}

\section{Gravitational waves from SGR superflares}

Another exciting feature of the last superflare is the possibility
of searching with good prospects for the first time an associated
gravitational wave (GW) burst. Technically, this kind of analysis
is akin to the search of associated bursts to the nearby GRB030329
\cite{LIGO}. One of the main advantages over "blind" searches is
that the time and position of the source (though not the relative
orientation of the emission pattern) are well-known, facilitating
the analysis.

The "mechanical" models have been explored as sources of GW in a
number of papers \cite{MVP,Pache}, and the general results seems
to be that the sharing of the released energy should favor the
ultimate dissipation as heat {\it via} radial oscillations
\cite{Sota} for slow rotators . However, since the the total
energy is huge, even a $1 \%$ or so being emitted in GW is still a
very large number and should suffice, in principle, for a
detection at the present sensitivities with high signal-to-noise
ratios. In these models, however, one finds difficult to
understand why the total energy budget (mainly carried by the
radiation) appears to be so reduced respect to the calculated
values, by factors ${O} \, \sim \, 10^{-6}$ or so. If, in turn,
the values of Table 1 for the {\it total} energy are adopted, then
the actual GW energy would be reduced accordingly, and the
prospects for its detection vanish.

To the best of our knowledge, no specific calculation of GW
emission has been performed for quaking solid stars as described
by Zhou et al. \cite{Xu2} as yet. The main reason for a higher
total energy release, perhaps up to $\Delta E \, \sim 10^{47} \,
erg$ can be attributed to a higher value of the shear modulus
$\mu$, which in turn allows to match the energy in gamma-rays
(Table 1) and also to expect quite naturally $\Delta E  \, \sim \,
E_{\gamma}$ on purely theoretical grounds. Thus, the model could
be considered to have a right energy scale and enough energy in GW
to be detected with the state-of-the-art experiments.

The GW expected from the propagation of a conversion $neutron \,
\rightarrow \, strange \, \, matter$ is strongly asymmetric in
presence of moderate magnetic fields, and therefore some GW signal
is expected \cite{Herman}, with a yet uncertain strength. The
details of this model of a superflare have been discussed in
Lugones et al.\cite{Bug}, as already noted. The overall prospects
seem encouraging for a positive detection. The same considerations
apply to the model presented in Mosquera Cuesta et al. \cite{MC},
even though the evidence for accretion-powered SGRs has not been
found so far \cite{TW}.

Ioka \cite{Ioka} has performed a detailed calculation of the
magnetospheric energy release model, with the result that the
changing deformation (increase in the moment of inertia) may
result in GW emission {\it via} the excitation of star
oscillations. The simplest estimates indicate about $10^{47} erg$
in GW, practically the same number than the exotic solid quake
model expectation. This stresses the need of further study of the
temporal and spectral {\it differences} predicted by both models.

On general grounds, one can resort to the simplest and roughest
estimate for the GW emission, largely independent of the model as
long as the energy release is quick enough (thus giving rise to a
burst of GW). The signal-to-noise ratio can be expressed in terms
of the basic quantity $E_{\gamma}$ if we write $\Delta E \, = \,
E_{\gamma}/\eta$. The quantity $\eta$ relating both energies could
be smaller than 1 (if heat generation is inefficient), or larger
than 1 (most of the energy coming out in gamma-rays), as suggested
above. The signal-to-noise ratio for a broadband interferometer
reads

\begin{equation}
{S\over{N}} \, =\, 10^{5} \eta^{-1/2} {\biggl( {E_{\gamma} \over
{M_{\odot} c^{2}}} \biggr)}^{1/2} {\biggl( {1 \, kHz \over
{f_{c}}} \biggr)}^{3/2} {\biggl({10 \, kpc \over {r}}
\biggr)}
\label{primera}
\end{equation}

where $f_{c} \, \sim \, 2 \, kHz$ is the characteristic frequency
of the signal, assumed to correspond to the lowest quadrupolar
mode of the star oscillation, and $r$ is the distance to the
source \cite{Hor}. Requiring $S/N \, \geq \, 3$ as a minimum
criterion for detection within a short time interval around the
superflare yields the limits on the parameter $\eta$ quoted in the
last column of Table 1 (the LIGO team uses a much conservative
criterion, $S/N \geq 8$ in their analysis). As is stands,the last
superflare from the source SGR 1806-20 is the only event which
could be detected even if the energy seen in gamma-rays carried
away most of the release.

\section{Conclusions}

We have briefly discussed some of the issues related to the
observed superflares from SGR sources. In particular, the
well-documented event of 2004 Dec 27 from the source SGR 1806-20
has restricted the number of viable models because of the involved
energetics \cite{Hur,Tera}.

Another relevant question is whether detectable GW emission
accompanies the gamma superflare. As it stands, the giant
superflare from SGR 1806-20 may be used to probe essentially all
available models in this respect, whereas the former events, in
addition of not being monitored, could only have probed models in
which the energy put in GW was overwhelming. Thus, a careful
search has to be performed in the data around $1 \, kHz$, this
frequency being dependent on the equation of state above nuclear
matter densities. Our conclusions are in line with other works,
including the latest \cite{Coc,Dub} which specifically discussed
the sources and detection strategies prior to the last recorded
event.

Note added in proof: after the acceptance of this work, we became
aware of a paper by Drago, Pagliara and Berezhani (gr-qc/0405145)
in which mini-collapses of rapidly spinning-down hybrid stars are
shown to produce GW bursts, a scenario perfectly compatible with
our statements in the general discussion.

\section{Acknowledgements}

We wish to acknowledge the guidance and advice of J.A. de Freitas
Pacheco and H. Mosquera Cuesta. The author wish to thank the S\~ao
Paulo State Agency FAPESP for financial support through grants and
fellowships, and the partial support of the CNPq (Brazil).

\end{document}